\documentclass[runningheads]{svmult}

\usepackage{makeidx}   % allows index generation
\usepackage{graphicx}  % standard LaTeX graphics tool
\usepackage{subeqnar}  % subnumbers individual equations
\usepackage{multicol}  % used for the two-column index
\usepackage{physprbb}  % modified textarea for proceedings,
\makeindex             % used for the subject index
\catcode`\@=11
\def\gsim{\ifmmode{\mathrel{\mathpalette\@versim>}}
    \else{$\mathrel{\mathpalette\@versim>}$}\fi}
\def\lsim{\ifmmode{\mathrel{\mathpalette\@versim<}}
    \else{$\mathrel{\mathpalette\@versim<}$}\fi}
\def\@versim#1#2{\lower 2.9truept \vbox{\baselineskip 0pt \lineskip 
    0.5truept \ialign{$\m@th#1\hfil##\hfil$\crcr#2\crcr\sim\crcr}}}
\catcode`\@=12
\begin{document}
\title*{A physically motivated toy model for the BH-spheroid coevolution}
\toctitle{A physically motivated toy model 
\protect\newline for the BH-spheroid coevolution}
% allows explicit linebreak for the table of content
%
\titlerunning{BH-spheroid coevolution}
% allows abbreviation of title, if the full title is too long
% to fit in the running head
%
\author{Luca Ciotti\inst{1}
\and Jeremiah P. Ostriker\inst{2,3}
\and Sergey Yu. Sazonov\inst{4,5}}
\authorrunning{Luca Ciotti et al.}
% if there are more than two authors,
% please abbreviate author list for running head
%
%
\institute{Dept. of Astronomy, Bologna University, Italy
\and  Institute of Astronomy, Cambridge, UK
\and Dept. of Astrophysical Sciences, Princeton University, USA
\and Max-Planck-Institut f\"ur Astrophysik, Garching bei M\"unchen, Germany
\and Space Research Institute, Russian Academy of Sciences, Moscow, Russia 
}

\maketitle              % typesets the title of the contribution

\begin{abstract}
We present a summary of the results obtained with a time-dependent,
one-zone toy model aimed at exploring the importance of radiative
feedback on the co-evolution of massive black holes (MBHs) at the
center of stellar spheroids and their stellar and gaseous
components. We consider cosmological infall of gas as well as the mass
and energy return for the evolving stellar population.  The AGN {\it
radiative} heating and cooling are described by assuming
photoionization equilibrium of a plasma interacting with the average
quasar SED.  Our results nicely support a new scenario in which the
AGN accretion phase characterized by a very short duty-cycle (and now
common in the Universe) is due to radiative feedback. The
establishment of this phase is recorded as a fossil in the Magorrian
and $M_{\rm BH} -\sigma$ relations.
\end{abstract}

\section{A simple scenario}

In [1,2,3] we discuss the importance of radiative feedback on the
co-evolution of MBHs at the center of spheroidal galaxies and that of
their stellar and gaseous components. After assessing the energetics
of the radiative feedback by using the results in [4], in [1,3] we
demonstrate that the observed relation between the MBH mass and the
stellar velocity dispersion can be the natural consequence of the MBHs
growth at early times in the galaxy history and the gas depletion of
the galaxy gaseous content due to star formation. In particular, in
[1] we show that \emph{radiative feedback from the central AGN is
effective in halting accretion} when the gas temperature is $> 3\times
10^4$ K, and the mass of the MBH has grown to a critical value:
\begin{equation}
{M_{\rm BH}^{\rm crit}\over 10^{11}M_{\odot}}\simeq \left(
                       {\sigma_*\over 200 {\rm km/s}}\right)^4
                       {L_{\rm Edd}\over L_{\rm BH}}
                       {M_{\rm gas}\over M_{\rm gal}}
\end{equation}
where $\sigma_*$ is the stellar velocity dispersion in the spheroid,
$M_{\rm gal}$ is its mass, $L_{\rm BH}$ is the accretion
luminosity. Under the assumption that $L_{\rm BH}\simeq L_{\rm Edd}$
during the phases of significant growth of the central MBH (e.g.,
[5,6,7]), one can derive a condition on $M_{\rm gas}/M_{\rm gal}$ in
order to satisfy the $M_{\rm BH}-\sigma$ relation observed today
(e.g., see [8,9,10]):
\begin{equation}
{M_{\rm gas}\over M_{\rm gal}}\simeq 10^{-3} 
            {L_{\rm BH}\over L_{\rm Edd}}.
\end{equation}
In [1,3], with the aid of a simple yet physically motivated toy
model, we show that such low-density, hot-temperature phases for the
ISM are established early in the lifetime of stellar spheroids as a
natural consequence of galaxy formation.

In our simulations we always find that at the epoch of the transition
between cold and hot solutions $10^{-3}< M_{\rm gas}/M_{\rm gal}<
10^{-2}$, in very good agreement with the request of
Eq. (2). Remarkably, if in our toy model we adopt duty-cycles of the
order of those obtained in numerical hydrodynamical simulations and
also derived from observations $(\sim 10^{-3}$, see [5,6,7]), we
obtain present-day MBHs that follow the observed scaling
relations. Substantially longer duty-cycles with (very) low accretion
efficiencies are not a viable alternative, because in this case
accretion would be dominated by Bondi accretion, resulting in too
massive MBHs.  Gas heating due to the evolving galaxy stellar
component in its various forms (stellar winds thermalization, SNII and
SNIa explosions), is confirmed to be an important ingredient in galaxy
evolution.  Due to their shallower potential well, low mass galaxies
host stronger galactic winds than more massive galaxies.  In the toy
model AGN heating is not sufficient ``per se'' to produce galaxy
degassing, even though is able to maintain the ISM temperature near to
the galaxy virial temperature.

Thus, a plausible (and observationally supported) scenario for the
coevolution of MBHs and stellar spheroids, is the following: at early
times in the galaxy life, when the protogalactic gas was dense and
cold, star formation and MBH growth were approximately parallel
[7]. Later on, due to the MBH growth on one side and the gas depletion
(due to star formation and reduced cosmological infall) on the other,
$M_{\rm BH}$ reaches the value $M_{\rm BH}^{\rm crit}$. After this,
accretion is characterized by a short duty cycle, of the order of
$10^{-3}$, with only a marginal growth of the central BH, even in
presence of the non negligible stellar mass losses of the passively
evolving stellar population [11].

%INDEX%%%%%%%%%%%%%%%%%%%%%%%%%%%%%%%%%%%%%%%%%%%%%%%%%%%%%%%%%%%%%%%
% Please check with the editor of your book whether he plans to
% include a "mutual" subject index - if so, please code your entries
% in the standard syntax. For your own purposes you may print your
% "personal" index by using the following commands:
%
%\clearpage
%\addcontentsline{toc}{section}{Index}
%\flushbottom
%\printindex
%%%%%%%%%%%%%%%%%%%%%%%%%%%%%%%%%%%%%%%%%%%%%%%%%%%%%%%%%%%%%%%%%%%%%

\end{document}